\documentclass[fleqn,12 pt]{article}
\usepackage[russian]{babel}
\usepackage{amssymb,amsmath}
\begin{document}
\newcommand{\E}{\mathcal{E}}

\title{\small\textbf{ QUANTUM OSCILLATOR IN THE THERMOSTAT AS A
MODEL IN THE THERMODYNAMICS OF OPEN QUANTUM SYSTEMS} }
\author{\small\textbf{A.D. Sukhanov}\footnote{This text was reported on the Bogoljubov conference at 4 September 2004, JINR, Dubna }}
\date {}
\maketitle
\begin{center} Bogolyubov's Laboratory of Theoretical Physics\\
JINR, Dubna, Moscow Region, Russia. Tel. (7-095)5012272. E-mail:
ogol@oldi.ru \end{center} \normalfont

\begin{abstract}
The quantum oscillator  in the thermostat is considered
as the model of an open quantum system. Our analysis will be heavily
founded on the use of the Schroedinger generalized uncertainties
relations (SUR).
Our first aim is to demonstrate that for the quantum oscillator the
state of thermal equilibrium belongs to the correlated coherent
states (CCS), which imply the saturation of SUR at any temperature.
The obtained results open the perspective for the search of some
statistical theory, which unifies the elements of quantum mechanics
and GDT; this in turn will give the foundation for the modification
of the standard thermodynamics.
\end{abstract}

In the last decades the interest to the thermodynamics of open
quantum systems raised greatly. To the number of these objects one
should, e.g., attribute nano-objects in the conditions of ultra-low
temperatures, the objects, revealing macroscopic quantum phenomena
and the effects of Casimir and Unruh as well as quark-gluon plasma,
models of early Universe and so on. Firstly, in these systems the
significant role play the fluctuations of conjugated physical
quantities and correlations between them. Further, there arises the
necessity of simultaneous account for the quantum and thermal
fluctuations, which are determined by the fundamental constants
$\hbar$ and $k_B$ .

However, no consistent theory for the description of such kind of
objects exists at the moment. For many years on this role was quite
reasonable pretended by the statistical mechanics Gibbs - von
Neumann and the thermodynamics founded on it. But this approach
appeared to be not always effective for the description of the open
quantum systems. First of all, it was not possible within this
approach to calculate accurately the Bekenstein - Hawking entropy
for the black holes. Finally it has brought to the pessimistic
conclusion: "The thermodynamics of the black holes gives some
weighty arguments in favor of the far-reaching non-adequacy even of
the generally accepted  understanding of the sense of matter's
"ordinary entropy" [1]. On the authoritative conference "Quantum
limits of the Second Law" [2] was quite seriously raised the problem
about the necessity of correction of thermodynamics laws -
especially of the second and third ones - in the direction of
significant account for the quantum effects. In particular, for the
long time there are considerable doubts in the validity of the Third
Law; the known form of this law follows  from the statistical
mechanics, according to which $S\rightarrow 0$          as
$T\rightarrow 0$ for any  objects.

The above arguments compel to search for the solution of problems of
this kind some replacement for the traditional statistical
mechanics. For many years in the physicist's public opinion there
was formed the firm conviction that this approach is almost unique
way of statistical description of nature; however, this is far from
being correct. Even Lorenz [3] in his lectures meant the statistical
theories in plural number; the same was written by Born [4] in
connection with Einstein's works.

Fenyes [5] was the first who showed convincingly that there exist
another statistical theories, which are principally different from
the statistical mechanics. On the one hand, this is the traditional
quantum mechanics and it's version in the form of the Nelson's
stochastic mechanics [6], on the other hand - the generalized
diffusion theory (GDT), founded on the Einstein - Fokker - Plank
equation [7-9]; the last theory includes the theory of Brownian
motion founded as long as Einstein [10]. At present time these
theories are different from one another; moreover, every of these
theories is, in turn, qualitatively different from the statistical
mechanics. On the other hand, these theories are conceptually
related to one another due to the significant role of fluctuations
(and their correlations) between the conjugated physical quantities.

In this report we consider the quantum oscillator in the thermostat
as the model of an open quantum system. Our analysis will be heavily
founded on the use of the Schroedinger generalized uncertainties
relations (SUR) [11], which are valid in both theories meant above.
Our first aim is to demonstrate that for the quantum oscillator the
state of thermal equilibrium belongs to the correlated coherent
states (CCS), which imply the saturation of SUR at any temperature.
The obtained results open the perspective for the search of some
statistical theory, which unifies the elements of quantum mechanics
and GDT; this in turn will give the foundation for the modification
of the standard thermodynamics.

In order to realize the fundamental cause of the qualitative
difference between the statistical mechanics and GDT it is
sufficient to consider the classical oscillator in the thermostat.
In this case the Maxwell - Boltzmann distribution function in the
phase space is of the form [12]:

\begin{equation}\label{1}
f_{M-B} (q,p) = \frac{1}{2\pi\Delta q \Delta p}\exp
{\{-\frac{q^2}{2\Delta q^2}-\frac{p^2}{2\Delta p^2}\}}=
f_B(q)f_M(p),
\end{equation}
where

\begin{equation}\label{2}
\Delta q^2 = \frac{k_{B}T }{m\omega_{0}^2};\;\;\; \Delta p^2 =
mk_BT.
\end{equation}

Due to the multiplicative form of the function $f_{M-B} (q,p)$ the
correlator of conjugated variables $q$           and $p$

\begin{equation}\label{3}
\sigma_{qp}\equiv \overline{(\Delta q \Delta p)}= 0.
\end{equation}
From the Eq. (3) it follows that the SUR [13] at $T\neq 0$ is of the
form of strong inequality

\begin{equation}\label {4}
\Delta p \Delta q \equiv \frac{k_{B} T }{\omega_0}> \sigma_{qp}=0.
\end{equation}.

At the same time due to Smoluchowski [14] (see also [15]) in GTD the
density of the distribution in coordinate, or configuration, space
(in the aperiodic regime at $t\gg \tau$                 ) equals

\begin{equation}\label {5}
\rho (q,t) = \frac{1}{\sqrt{2\pi \Delta q^2}}\exp\{
-\frac{q^2}{2\Delta q^2}\},
\end{equation}
where

\begin{equation}\label {6}
\Delta q^2 = \frac{D_{dif}}{\omega_0} (1-\exp\{-2Q^2{t/\tau}\});
\;\;\; Q=\omega{_0}\tau; \;\; 0\leqslant Q\leqslant 1/2;
\end{equation}

\begin{equation}\label {7}
D_{dif} = \frac{k_B}{m\omega_0}.
\end{equation}.

As is well known, in the theory of Brownian motion the definition of
the moment of the individual particle $p=m\dot{q}$ , accepted in
statistical mechanics, is invalid. Thus following Furth [16] and
Fenyes [5] as a moment conjugated to the random coordinate $q$ is by
definition accepted the moment of the particles' flow (at
$v_{macro}=0$ )

\begin{equation}\label {8}
p\equiv mv_{dif}= m\frac{j_{dif}}{\rho}
=\frac{m}{\rho}(-D_{dif}\frac{\partial\rho}{\partial
q})=mD_{dif}\frac{q}{\Delta q^2}.
\end{equation}.

To the state of the thermal equilibrium there corresponds the limit
$t\rightarrow \infty$           at $Q\neq 0$             , when

\begin{equation}\label {9}
\Delta q^2 =\frac{D_{dif}}{\omega_0}.
\end{equation}
In the same state

\begin{equation}\label {10}
\Delta p^2= (mD_{dif})^2 \frac{\Delta q^2}{(\Delta q^2)^2} =
m^2\omega_0 D_{dif};
\end{equation}

\begin{equation}\label {11}
\sigma_{qp}= \overline{\Delta q \Delta p}= m\int dq\;\;\; q
j_{dif}(q) = mD_{dif}.
\end{equation}
It follows from here that SUR in GTD are of the form of equality

\begin{equation}\label {12}
\Delta p \Delta q \equiv \frac{k_BT}{\omega_0} = \sigma_{qp}\equiv
mD_{dif} = \frac{k_B T}{\omega_0}.
\end{equation}

This denotes that the state, described by the function $\rho(q,t)$
of the form (5) at $Q\neq 0$ and $t\rightarrow\infty$ is the CCS
state. Note that for the free Brownian particle when $\omega_{0}=0,$
one obtains the analogous result, but $\omega_{0}$ should be
replaced by $1/\tau$ [16].

It is not difficult to note that the expressions for $\Delta q^2$
and $\Delta p^2$ in the state of thermal equilibrium (and hence the
left-hand sides of SUR in both statistical theories considered
above) coincide. Accordingly coincide in form also the limiting
function $\rho (q)$                  and it's Fourier-transform
$\rho (p)$ with the functions $f_B(q)$ and $f_M(p)$ . However the
expressions for the correlators $\sigma_{qp}$ and thus the
right-hand sides of SUR in them are qualitatively different.

From these arguments there follow some important conclusions:\\
-statistical mechanics differs from GTD in the definition of the
conjugated moment. In the former theory $p$                  and $q$
are independent, whereas in the latter - dependent ones;\\ - the
cancellation of the quantity $\sigma_{qp}$                 in the
statistical mechanics means that in this theory the diffusion is
fully neglected ($D_{dif}=0$               );\\ - in turn, this
means that in the statistical mechanics there is not accounted for
the diffusion flow $j_{dif}$ , caused by the non-uniform coordinate
distribution which is brought about due to presence of the
fluctuations ($\Delta q^2\neq0$         ).

In other words, the statistical mechanics doesn't give the
consecutive description of the Nature, where three second-order
moments $\Delta p^2$                    , $\Delta q^2$ and
$\sigma_{qp}$              would be considered at the same rights.

In order to more convenient comparison with the classical variants
of the theories let us use as the version of quantum mechanics the
Nelson's stochastic mechanics; accordingly, in quantum statistics we
will go out from the Wigner function [17]. According to Nelson, with
the wave function in coordinate representation $\phi(q) =\sqrt{\rho
(q)}\exp \{i\varphi(q)\}$ one may associate two velocities defined
by the gradients of the phase $\varphi(q)$ and density $\rho(q)$ in
the   following way:

\begin{equation}\label {13}
v_{qu}= \frac{\hbar}{m}\frac{\partial \varphi}{\partial q} =2
D_{qu}\frac{\partial \varphi}{\partial q}; \;\;\; u_{qu}=
-D_{qu}\frac{1}{\rho}\frac{\partial \rho}{\partial q},
\end{equation}
where $D_{qu}\equiv\frac{\hbar}{2m}$                        is
usually called quantum diffusion coefficient. From these two
quantities just the $v_{qu}$ enters the continuity equation, so
physically this quantity is related to the velocity $v_{dif}$ in
GDT. Taken formally, the velocity $v_{dif}$                    is
more similar to the quantity $u_{qu}$ , but actually the former is a
truly quantum quantity, connected with the non-commutativity of the
operators $\widehat{p}$ and $\widetilde{q}$                     .

Further analysis is founded on the fact that as a conjugated moment
is introduced (by definition) [13] the complex quantity
$\widetilde{P}$ , which combines the contributions of both
velocities of the form (13)
\begin{equation}\label{14}
\widetilde{P}\equiv m(v_{qu}+ i u_{qu}).
\end{equation}
Then
\begin{equation}\label{15}
\Delta p^2 = \overline{|\Delta \widetilde{P}^2|} =
\overline{(mv_{qu})^2} + \overline{(mu_{qu})^2};
\end{equation}

\begin{equation}\label{ 16}
\left[\;\;{\overline{\mid\widetilde{R}_{qp}\mid}}\;\;\right]^2 =
\sigma_{qp}^2 + c_{qp}^2 = m^2 \left\{[\overline{qv_{qu}}]^2
+[\overline{qu_{qu}}]^2 \right\}.
\end{equation}

Here $\sigma_{qp}$                     and $c_{qp}$ are the
contributions of anticommutator and of commutator of the operators
$\widehat{q}$                         and $\widehat{p}$ accordingly
(in the scopes of traditional quantum mechanics), whereas
$\widetilde{R}_{qp}$ is the generalized complex correlator entering
the right-hand side of SUR [13].

For the quantum oscillator in CCS the wave function is of the form
[13]

\begin{equation}\label {17}
\psi(q) = \frac{1}{{(2\pi \Delta q^2)}^{1/4}} \exp\{
{-\frac{q^2}{4\Delta q^2}}(1-i\alpha)\},
\end{equation}
where $\alpha$                    is the parameter fixing the
particular CCS. In accordance with the Eqs. (14) and (13) it holds
for this state
\begin{equation}\label{18}
\widetilde{P} = \frac{mq}{\Delta q^2} (D_{qu}\alpha)
+i\frac{mq}{\Delta q^2} D_{qu}.
\end{equation}
For the oscillator the contribution into $v_{qu}$ from the motion of
the wave packet is absent; correspondingly,

\begin{equation}\label {19}
\left[\;\;{\overline{\mid\widetilde{R}_{qp}\mid}}\;\;\right]^2 = m^2
[(D_{qu}\alpha)^2 + D_{qu}^2].
\end{equation}
Using further Eqs. (15) and (18), we obtain as a SUR the equality;
indeed, as was awaited for CCS, the given form of SUR is saturated:

\begin{equation}\label {20}
\Delta p^2 \Delta q^2 =
\left[\;\;{\overline{\mid\widetilde{R}_{qp}\mid}}\;\;\right]^2 = m^2
D_{qu}^2 ({\alpha}^2 +1).
\end{equation}
Recall, that in the ground state (when CCS goes over into CS) the
parameter $\alpha=0$             , so that

\begin{gather}\label{ 21}
\Delta q^2 = \frac{D_{qu}}{\omega_0};\;\; \Delta p^2 = m^2 \omega_0
D_{qu};\;\; \nonumber
\\\sigma_{qp}= 0 ;\;\;\overline{|\widetilde{R}_{qp}|}= c_{qp}= mD_{qu}=
\frac{\hbar}{2}.
\end {gather}

Thus, as concerns the form of SUR, the oscillator in quantum
mechanics appears to be analogous on the classical oscillator in GDT
when $t\gg\tau$                         . All the difference
consists in the fact that the right-hand side of SUR in the second
case is ensured exclusively on account of $\sigma_{qp}$ whereas in
the second case this goal is achieved only by means of both
contributions -- $\sigma_{qp}$ as well as $c_{qp}$ . In fact, this
signifies that in quantum mechanics as well as in GDT the conjugated
quantities -- coordinate and momentum -- are mutually dependent. In
this case the difference is only that this dependence in terms
$\sigma_{qp}$ appears always in explicit form. On the other hand, in
the term $c_{qp}$ in the traditional formulation of quantum
mechanics this mutual dependence is implicit: it becomes explicit
only in the framework of the statistical mechanics.

The analysis carried above has shown that the accepted definitions
of the conjugated momenta lead to the following results:
\par\bigskip\noindent
\begin{tabular}{|c|c|c|}
\hline Classical GDT  & Quantum Mech. in CS   & Quantum Mech. in
CCS\tabularnewline $(T\gg 0)$ & $(T= 0)$ &$(T\geqslant 0)$\\
\hline   $\Delta p^2 \Delta q^2 = (mD_{dif})^2$ & $\Delta p^2 \Delta
q^2 = (mD_{qu})^2$ & $\Delta p^2 \Delta q^2 = (mD_{qu})^2 (\alpha^2
+1)$\\ \hline
\end{tabular}
\par\bigskip
Thus we obtain two limiting cases ($T\rightarrow 0$ and high $T$)
and also the structure of the general expression for the right-hand
side of SUR for the oscillator in CCS. The problem is to find such
an expression for the parameter $\alpha$  which would correspond to
the quantum oscillator in the state of thermal equilibrium at any
values of $T$                   .

In order to solve this problem we use the fluctuation - dissipation
theorem (FDT) following from most basic principles of quantum and
statistical physics [12]. Applying this theorem to the coordinate
fluctuations we obtain for it's dispersion the following expression:
\begin{equation}\label{22}
\Delta q^2 = \frac{1}{2\pi} \int \limits_{-\infty}^\infty
\alpha^{''}(\omega)\left(\frac{\hbar}{2}\coth
\frac{\hbar\omega}{2k_BT}\right) d\omega.
\end{equation}

Let us emphasize that the function $\coth\frac{\hbar\omega}{2K_BT}$
entering Eq. (22) does not only belong to any specific oscillator
but is a universal function for any systems and, moreover, for
fluctuations of any physical quantities in these systems. All the
specific information about the concrete system is contained in the
imaginary part of the generalized susceptibility
$\alpha^{''}(\omega)$.

Let us define the generalized diffusion coefficient as follows:
\begin{equation}\label{ 23}
D_{gen}\equiv D_{qu} \coth\frac{D_{qu}}{D_{dif}} = \sqrt{D^2_{term}+
D^2_{qu}}.
\end{equation}
This quantity characterizises the diffusion flow due to
the non-uniformity in the coordinate distribution, which
arises on account of the quantum as well as the thermal effects. Here
\begin{equation}\label{ 24}
D_{term} \equiv \frac{D_{qu}}{\sinh\frac{D_{qu}}{D_{dif}}}
\end{equation}
is the thermal part of the generalized coefficient $D_{gen}$ ,
depending both upon $\hbar$ and $k_B$               , while at
$T\rightarrow 0$ the coefficient $D_{term}\rightarrow 0$ , whereas
at high $T$ the coefficient $D_{term}\rightarrow D_{dif}$ has the
form (7).
On account of the definition (23) FDT for the fluctuation
of the coordinate may be presented in the form
\begin{equation}\label{ 25}
\Delta q^2 =
\frac{1}{2\pi}\int\limits_{-\infty}^{\infty}\alpha^{''}(\omega)m
D_{gen}(\omega) d\omega
\end{equation}
which emphasizes the rigid connection between the characteristic of
fluctuation $\Delta q^2$  and the generalized diffusion coefficient
$D_{gen}$                               .
For the oscillator with
weak damping
\begin{equation}\label{ 26}
\alpha^{''} = \frac{1}{|\widetilde{\gamma}|} \delta (\omega -
\omega_{eff}) = \frac{1}{m\omega}\delta(\omega-\omega_{eff})
\end{equation}
where $\omega_{eff} = \sqrt{\omega_{0}^2 + 1/\tau^2}\approx
\omega_{0}$ , so as $\Delta q^2 = D_{gen}/\omega_{0}$ what obviously
generalizes Eqs. (2) and (9). Note that in reverse limit $1/\tau\gg
\omega_{0}$ from Eqs. (25) and (26) follows the well known Einstein
relation for the free Brownian particle

The problem of quantum oscillator in the framework of statistical
mechanics was firstly solved by Bloch [18]; later on the original
solutions of this problem were also found by Landau and Lifshitz
[12], Leontovich [19] and Feynman [20]. In sake of simplicity let us
use the Wigner function obtained on he grounds of the Gibbs' density
operator:
\begin{equation}\label{ 27}
\widehat{\rho}_G =  \frac{\exp \{-\widehat{H}/k_BT\}}{Sp\;\left
(\exp \{-\widehat{H}/k_BT\}\right)}
\end{equation}
with the operator $\widehat{H}= \frac{\widehat{p}^2}{2m}+
\frac{m\omega_{0}^2q^2}{2}$ for the case of oscillator
\begin{equation}\label{ 28}
W(q,p) = \frac{1}{2\pi \Delta q \Delta q} \exp{\{-\frac{q^2}{2\Delta
q^2} - \frac{p^2}{2\Delta p^2}\}},
\end{equation}
where
\begin{equation}\label{ 29}
\Delta q^2 = \frac{D_{gen}}{\omega_0};\;\;\;\; \Delta p^2 = m^2
\omega_0 D_{gen}.
\end{equation}
Note that the expression (28) reminds in form the Maxwell -
Boltzmann distribution with the substitution of $D_{dif}$ for
$D_{gen}$ in Eq. (29). In analogy with the Eq. (1) from the Eq. (28)
it follows that
\begin{equation}\label{ 30}
\rho (q) = \frac{1}{\sqrt{2\pi\Delta
q^2}}\exp\left\{-\frac{q^2}{2\Delta q^2}\right\};\;\;\rho (p) =
\frac{1}{\sqrt{2\pi\Delta p^2}}\exp\left\{-\frac{p^2}{2\Delta
p^2}\right\}.
\end{equation}
As is well known, the limiting expressions for these quantities at
$T\rightarrow 0$                   or $T\gg 0$ go over accordingly
into the probability densities of oscillator's ground state in $q$-
or $p$- representations, or in the Boltzmann $f_{B}(q)$ or Maxwell
$f_{M}(p)$ distribution functions.

It is easily seen that due to the multiplicativity of the Wigner
function in this case one has $\sigma_{qp}=0$ . That's why at the
arbitrary temperature in the framework of quantum statistical
mechanics SUR for the oscillator obtains the form
\begin{equation}\label{ 31}
\Delta p^2 \Delta q^2 \equiv (mD_{gen})^2 \geqslant
\left[\;\;{\overline{\mid\widetilde{R}_{qp}\mid}}\;\;\right]^2 =
c^2_{qp} = \frac{\hbar^2}{4}.
\end{equation}
The equality sign is only achieved here  at $T=0$ ; in the case of
$T\neq 0$ just as in classical statistical mechanics SUR (31) has
the form of the strong inequality
\begin{equation}\label{ 32}
\Delta p \Delta q > \frac{\hbar}{2}.
\end{equation}
This means that for the oscillator in the thermostat the single
distinction in the descriptions in the frameworks both of the
quantum and the classical statistical mechanics is connected with
the account for the hidden mutual dependence  between the conjugated
coordinate and momenta. This dependence is brought about by the
non-commutativity of the operators of corresponding variables which
lead to the presence of the quantum fluctuations $\Delta q^2$ ,
characterizing the oscillator's ground state. However, the account
for the mutually depending thermal and quantum fluctuations of the
coordinate, which take place outside of the ground state, is absent
both in quantum and classical forms of statistical mechanics.

The arguments presented above lead to the conclusion, that for the
adequate description of the quantum oscillator in the thermostat the
quantum statistical mechanics is to a large extent invalid. One
should choice another variant of the theory, which allows to account
simultaneously for quantum and thermal effects and uses for this
goal the earlier  obtained generalized diffusion coefficient
$D_{gen}$  of the form (23). It seems that arising difficulties are
the result of the non-consequent account for the influence of the
thermostat on the oscillator, namely of the infinite degrees of
freedom by the thermostat.

This remark make us arrive to the notion to go over to the
description of the considered open quantum system in the framework
of the quantum field theory at finite temperatures (this theory is
now rather actively used in many applications).

There exist two most elaborated versions of this theory:\\- -   the
version, founded on the temperature Green functions and the
imaginary time (Matsubara [21]);\\- -   the version, founded on the
Thermo Field Dynamics (TFD) with the real time (Araki [22], Umezawa
[23, 24]).

The first of these versions was especially popular thus far;
however, it is not suited for the use of causal Green functions
which form the foundation of the quantum field theory at $T=0$, what
prevents the uniform description of physical phenomena at arbitrary
temperatures. Besides this formal circumstance the TFD is of
interest because it is largely founded on the three fundamental
ideas of Bogoluybov - namely, his $(u,v)$-transformation, method of
quasiaverages and the description of the spontaneous violation of
symmetry.

The main physical idea of TFD consists in the following: the placing
of the system in the thermostat (or, more precisely, bringing them
in the state of the thermal contact with one another) is effectively
equivalent to the doubling of the system's degrees of freedom, which
in turn leads to the removal of the degeneration of the system. The
influence of the thermostat manifests itself in the fact that
instead of one way there arise effectively two independent ways of
the absorbtion of the external energy by the system.

These two ways are the following ones: the first is the ordinary
appearance of the excitation of new quanta of energy, while the
second is the disappearance of the vacancies, which were "induced"
in the system's spectrum under the influence of the thermostat at
$T\neq0$. In this case both annihilation $ \widetilde{a}$ and
creation $\widetilde{a}^+$ operators perform two independent
operations (instead of one at $T=0$), so these operators may be
represented by the linear combination of another operators $ b_i,
b_j^+ (i,j=1,2)$:
\begin{equation}\label{ 33}
\widetilde{a}= ub_1 +vb_2^+; \;\;\;\; \widetilde{a}^{+}= u^*b_1^+ +
v^* b_2^+
\end{equation}
with $\mid u\mid^2 - \mid v\mid^2 = 1$              . This
combination is the typical Bogoluybov $(u,v)$-transformation, where
\begin{equation}\label{ 34}
u=\cosh \tau \exp (i\varphi);\;\;\; v=\sinh\tau\exp(-i\varphi)
\end{equation}
at $0\leqslant \tau<\infty, -\pi \leqslant \varphi \leqslant\pi$

The operators $b_{i}, b^+_{i}$               characterize the
original particles and act in Fock space with the vacuum $\mid 0 >$
. The operators $\widetilde{a}^+$         and $\widetilde{a}$
characterize the quasiparticles and act in another Fock space  with
the vacuum $|0\rangle\rangle$ which is temperature dependent and
possess the property $\widetilde{a}|0\rangle\rangle=0$ . The vector
$|0\rangle\rangle$ is orthogonal to all vectors of the original Fock
space and so it doesn't belong to this space. In other words, the
transformation (33) is the canonical one, but in this case it leads
to the unitary non-equivalent representation of the canonical
commutation rules (CCR), because the thermostat is the system with
infinitely many degrees of freedom.

The main axiom of TFD comes in fact from the idea of the method of
quasiaverages: there is suggested that the original ground state
$|0\rangle$ is unstable relative to the perturbation created by the
thermostat. That's why it is assumed that the average of the number
operator of quasiparticles $\widetilde{a}^+\widetilde{a}$ at the
temperature $T\neq0$ over the original Fock vacuum $|0\rangle$ is
equal to the average of the number operator of original particles
$a^+a$ at $T=0$ over the Gibbs distribution with temperature
$T\neq0$ , i.e. calculated in the framework of quantum statistical
mechanics:
\begin{equation}\label{ 35}
\langle 0|\widetilde{a}^+\widetilde{a}|0\rangle \equiv
Sp\;\;(a^+\;a\;\widehat{\rho}_G)
\end{equation}
where the density operator $\widehat{\rho}_{G}$ depends upon
$\widehat{H}=\hbar\omega_{0}(a^+a+1/2)$ .

From the axiom meant above it is not difficult using the Eqs. (33)
and (34) to obtain the relation between the parameters of Bogoluybov
transformation and the temperature of the thermostat:
\begin{equation}\label{ 36}
\sinh^2\tau = \frac{1}{\exp
{\frac{\hbar\omega_0}{k_BT}}-1}=\frac{1}{2}\coth
\frac{D_{gen}}{D_{dif}}-\frac{1}{2}
\end{equation}
and analogously for $\cosh^2 \tau$             .

The obtained relation (36) allows to fix ultimately the quantity
$\alpha$ , which defines the phase of the oscillator wave function
in CCS in the presence of thermostat. If one puts $\varphi= \pi/4$
in the general expressions for $\Delta q^2$             and $\Delta
p^2$ and then use the Eqs. (36) and (23), one obtains
\begin{equation}\label{37}
\Delta q^2 = \frac{l_0^2}{2}\cosh 2\tau =
\frac{D_{gen}}{\omega_0};\;\Delta p^2 = m^2 \omega_0 D_{gen};
\end{equation}

\begin{equation}\label{ 38}
\alpha = \sinh 2\tau = \frac{1}{\sinh\frac{D_{qu}}{D_{dif}}};
\sigma_{qp} =mD_{qu}\alpha =
\frac{mD_{qu}}{\sinh\frac{D_{qu}}{D_{dif}}} = m D_{term}.
\end{equation}

It follows from these relations that for the oscillator in TFD the
SUR possess the form of equality

\begin{equation}\label {39}
\Delta p^2 \Delta q^2 = \sigma_{qp}^2 + c_{qp}^2 = (mD_{term})^2 +
(mD_{qu})^2 =(mD_{gen})^2,
\end{equation}
what naturally corresponds to the limiting cases (12) and (20) (at
the condition $\alpha=0$                      ).

It can also be shown that Wigner function in this case is equal to
the following expression:

\begin{equation}\label {40}
W(q,p)=\frac{1}{\pi\hbar}\exp\left\{ - \left[\frac{q^2}{2\Delta q^2
}+\frac{p^2}{2\Delta p^2}\right ]\coth\frac{D_{qu}}{D_{dif}}
-\frac{qp}{\Delta q\Delta p}\;\;
\frac{1}{\sinh\frac{D_{qu}}{D_{dif}}}\right\}.
\end{equation}
This function is not multiplicative in the arguments $q$ and $p$
what reveals the mutual statistical dependence of these arguments,
and it is this dependence which leads to the correlator
$\sigma_{qp}$ . Exactly the same result may be obtained [24] in
another way, namely by the direct account for the interrelation
between the operators $b_1$ and $b_2$            , which arise due
to the following fact: the quasiparticle vacuum $|0\rangle\rangle$
contains the condensate consisting from the pairs of particles.

Let us formulate some conclusions. \\ 1.  SUR (38) may be presented
in the following form:
\begin{equation}\label{ 41}
\Delta p \Delta q = \frac{\hbar}{2}\coth \frac{\hbar
\omega_0}{2k_BT} \equiv \frac{\hbar^*}{2}.
\end{equation}
In other words, for the oscillator in the thermostat the role of the
effective quantum of action plays the quantity $\hbar^*\geqslant
\hbar$ , which grows with growing temperature. The limiting values
of this quantity are $\hbar$                  (at $T\rightarrow 0$ )
and $2k_BT/\omega_{0}$ (at $T\gg0$).

2. The phase of the wave function for CCS of the oscillator in the
thermostat depends upon the temperature
\begin{equation}\label{ 42}
\varphi_{dif} = \varphi - \varphi_0 = \frac{q^2}{4\Delta q^2}\alpha
= \frac{q^2}{2l_0^2}\;\;\frac{1}{\cosh \frac{D_{qu}}{{D_{dif}}}}.
\end{equation}

At $T\rightarrow 0$                  the phase $\varphi\rightarrow
0$ ($ \sigma_{qp}\rightarrow 0$ ), at $T\gg 0$ the phase
$\varphi_{dif}\rightarrow q^2/2l_0^2\neq 0$ ($\sigma_{qp}\rightarrow
mD_{dif}=k_BT/\omega_0$ ). The last relation means that the role of
the phase of the wave function is preserved also in traditionally
classical region thus ensuring the existing of the thermal diffusion
in the classical limit.

3. In the given case CCS is characterized by the wave function and
so this state is a pure one; on the other hand, the parameters of
this wave function and the corresponding correlator $\sigma_{qp}$
depend upon the temperature and thus reveal the presence of the
thermal noise in this state [24]. The calculation of the entropy
(defined according to von Neumann) for the oscillator in CCS gives
$S=0$      at any temperature. This contradiction makes very actual
the question of the more correct definition of the entropy, which
should be able to account for the thermal noise in CCS.

4. For the subsequent generalization of thermodynamics it is useful
to introduce the notion of the generalized temperature [25]. Let us
define the following temperatures
\begin{equation}\label{ 43}
T_{gen}\equiv T_{qu}\coth\frac{T_{qu}}{T};\;\;
T_{qu}\equiv\frac{\hbar\omega}{2k_B}.
\end{equation}

Evidently, at high temperatures $T_{gen}\rightarrow T$ , but at
$T\rightarrow 0$ we have $T_{gen}\rightarrow T_{qu} \neq 0$, what
corresponds to the generalization of the Third Law. Note that
$T_{qu}$ is the characteristic of those mode of normal vacuum
vibrations which is in the resonance to the considered oscillator.

5. It is interesting to compare SUR "coordinate - momentum" with the
earlier obtained SUR "energy - frequency" and "energy - temperature"
for the oscillator in CCS in the thermostat, which are expressed by
the aid of the generalized temperature [26]
\begin{equation}\label{ 44}
\Delta p \Delta q = \frac{\hbar}{2} \frac{T_{gen}}{T_{qu}} =
\frac{k_B}{\omega}T_{gen};
\end{equation}

\begin{equation}\label{ 45}
\Delta \; \E \Delta \left(\frac{1}{T_{gen}}\right)= k_B.
\end{equation}

Note that in Eq. (44) the quantities $T_{gen}$       and $T_{qu}$
relate to the surrounding (i.e., the thermostat) and so don't
fluctuate. On the contrary, in Eq. (45) all quantities relate to the
system and so do fluctuate.

6. The comparison of the descriptions of the oscillator in the
thermostat in the frameworks of the quantum statistical mechanics
and TFD reveals that the expressions for $\rho(q)$ , $\rho(p)$     ,
$\Delta q^2$ and $\Delta p^2$ in these theories coincide completely.
However, the Wigner function and correlators $\sigma_{qp}$  in these
theories differ due to the difference of the definitions of the
conjugate momentum; it is just this difference which leads to
qualitatively forms of SUR. Note that this difference is preserved
also in quasiclassical limit ($\hbar\rightarrow 0$ ), what is
important for the theory of Brownian motion. Namely, it denotes that
the traditional opinion about the equivalence of the variants of
this theory - i.e., based on the Langevin equation (statistical
mechanics) and based on the Einstein - Fokker - Planck equation
(GDT) - needs some corrections.

7. The analysis carried above has also shown that the further
development of TFD [23, 24] is the most perspective direction of
search for the effective generalization of thermodynamics on the
quantum systems at ultralow temperatures, where the role of both
quantum and thermal fluctuations becomes significant. There is
urgent need for the more consistent spreading of the seminal ideas
of the quantum field theory on various thermal phenomena and ideas
of the thermodynamics on any quantum open systems. It can be hoped
that  this way may just lead to the complete realization of the
concept of the physics as a "single whole" science.

\begin{enumerate}

\item  Bekenstein J.D. Quantum Black Holes as Atoms. gr-qc/9901033
v.1
\item  Quantum Limits of Second Law. Ed. D.P. Sheehan, AIP,
San-Diego, 2002
\item  Lorentz H.A. Les Theories Statistiques en Thermodinamique.
 Leipzig-Berlin, Teubner, 1916.

\item  Born M. Einstein's statistical theories. Albert Einstein: Philisopher-Scientist, ed. P.A. Schilpp, N. Y., Tudor, 1949, p. 163

\item  F\'{e}nyes I. Zs. Phys. 132, 81, 1952

\item  Nelson E. Phys.Rev. 150, 1079, 1966;
Dynamical Theories of Brownian  Motion.
 Princeton. Univ Press, Princeton, 1967
\item  Einstein A. Ann. Phys. 19, 371, 1906.

\item  Fokker A.D. Ann. Phys., 43, 810, 1914;
\item  Plank M. Sitzungsber. Akad. Wiss., 324, 1917
\item  Einstein A. Ann. Phys. 17, 549, 1905.
\item  Schr\"{o}dinger. Ber. Kgl. Akad. Wiss. Berlin, 1930, S. 296

\item  Landau L.D. Lifshitz E.M. Statistical Physics P. I. Oxford,
Pergamon (1980)

\item  Sukhanov  À.D. Theor. and Math. Physics, 132(3), 1277,2002
\item  Smoluchowski Ì.  Bull. Intern de l'Ac. de Sciences de
Cracovie. (A), 418, 1913
\item  Ansel'm À.I. Foundations of Statistical Physics and Thermodynamics (in Russian) M., Nauka, 1973.
Ì., Íàóêà, 1973
\item  Furth R. Zs. Phys. 81, 143, 1933
\item  Wigner E. Phys.Rev. 40,749 (1932)
\item  Bloch F. Z.Phys. 74,295 (1932)
\item  Leontovitch Ì.À. Introduction to  Thermodynamics.
Statistical Physics (in Russian) Ì., Nauka. Physmathlit, 1983
\item  Feinman R.Statistical Mechanics. Ìassachusetts, Benjamin Inc. 1972
\item  Matsubara T. Prog. Thor. Phys. 14, 351 (1955)
\item  Shwinger J.  J.Math. Phys. 2, 407 (1961); Êåëäûø Ë.Â. ÆÝÒÔ,
47, 9, 1515 (1964)
\item  Umezawa H, Matsumoto H. Tachiki M. Thermo Field Dynamics
and Condensed States. North-Holl. Publ. Amsterdam, 1982
\item  Umezava H. Physica A. 158, 1, 306 (1989); Advanced Field
Theory Micro, Macro and Thermal Physics. N.Y. AIP 1993
\item  Sukhanov A.D. On the  Global Interrelation between Quantum
Dynamics and Thermodynamics. Proc. of XI Intern. Conf. "Problems of
Quantum Field Theory". 1998. Dubna, 1999. P. 232
\item  Sukhanov À.D. Theor. and Math. Physics, 125, (2), 1489, 2000;

\end{enumerate}

\end{document}